\documentclass[%
 reprint,
 amsmath,amssymb,
 aps,nofootinbib,
onecolumn]{revtex4-2}
\usepackage{tikz}
\usepackage{xcolor}
\usepackage{multirow}
\usepackage{graphicx}% Include figure files
\usepackage{dcolumn}% Align table columns on decimal point
\usepackage{bm}% bold math
\usepackage{fancyhdr}
\usepackage[colorlinks]{hyperref}
\usepackage[nameinlink,noabbrev]{cleveref}
\hypersetup{
    colorlinks=true,
    linkcolor=blue,
    filecolor=magenta,      
    urlcolor=blue,
   citecolor=blue
}
\usepackage{graphicx}% Include figure files
\usepackage{dcolumn}% Align table columns on decimal point
\usepackage{bm}% bold math
\usepackage{scalerel}
\usepackage{tikz}
\usetikzlibrary{svg.path}

\definecolor{orcidlogocol}{HTML}{A6CE39}
\tikzset{
  orcidlogo/.pic={
    \fill[orcidlogocol] svg{M256,128c0,70.7-57.3,128-128,128C57.3,256,0,198.7,0,128C0,57.3,57.3,0,128,0C198.7,0,256,57.3,256,128z};
    \fill[white] svg{M86.3,186.2H70.9V79.1h15.4v48.4V186.2z}
                 svg{M108.9,79.1h41.6c39.6,0,57,28.3,57,53.6c0,27.5-21.5,53.6-56.8,53.6h-41.8V79.1z M124.3,172.4h24.5c34.9,0,42.9-26.5,42.9-39.7c0-21.5-13.7-39.7-43.7-39.7h-23.7V172.4z}
                 svg{M88.7,56.8c0,5.5-4.5,10.1-10.1,10.1c-5.6,0-10.1-4.6-10.1-10.1c0-5.6,4.5-10.1,10.1-10.1C84.2,46.7,88.7,51.3,88.7,56.8z};
  }
}

\newcommand\orcidicon[1]{\href{https://orcid.org/#1}{\mbox{\scalerel*{
\begin{tikzpicture}[yscale=-1,transform shape]
\pic{orcidlogo};
\end{tikzpicture}
}{|}}}}
\usepackage{amsmath}
\usepackage{amssymb, amsfonts}
\usepackage[greek,english]{babel}
\begin{document}
\preprint{APS/123-QED}

\title{Jeans analysis in modified gravity: A general formulation}% Force line breaks with \\
%\thanks{Essay written for the Gravity Research Foundation 2023 Awards for Essays on Gravitation.}%

\author{Ryan Khaled}
%\email{hakim9502.benkrane@gmail.com, abdelhakim.benkrane@univ-ouargla.dz
%}
\address{Theoretical Physics Laboratory, Faculty of Physics, University of Bab-Ezzouar, USTHB, Boite Postale 32, El Alia, Algiers 16111, Algeria}

\author{Kamel Ourabah \orcidicon{0000-0003-0515-6728}}\email{kam.ourabah@gmail.com, kourabah@usthb.dz}
\address{Theoretical Physics Laboratory, Faculty of Physics, University of Bab-Ezzouar, USTHB, Boite Postale 32, El Alia, Algiers 16111, Algeria}

\date{\today}% It is always \today, today,

\begin{abstract}
We study the Jeans instability of a self-gravitating fluid in the context of alternative gravitational theories. Our formulation accommodates arbitrary corrections to the Newtonian potential, providing a versatile approach to analyzing a wide range of gravity theories in the weak-field limit. As concrete examples, we explore in detail three specific modifications: the Maneff potential, a logarithmic correction, and a Yukawa-type correction to Newtonian gravity. These three cases serve as representative examples of the range of functional forms—power-law, logarithmic, and exponential—that one might expect to arise in alternative theories of gravity. We determine the conditions for instability and the growth rate of perturbations for these potentials. To evaluate their validity, we compare the established stability criteria with the observed stability data from Bok globules, thereby enabling us to place constraints on the parameters of these alternative gravitational theories. Additionally, we extend our analysis to the quantum regime, offering a broader perspective on the interplay between gravity and quantum effects in the context of alternative theories of gravity.

\end{abstract}

\maketitle

%\tableofcontents
\section{Introduction}

Newtonian gravity provides the foundation for understanding classical gravitational phenomena but notoriously encounters significant limitations in certain astrophysical contexts \cite{1}. General relativity (GR) has addressed many of these shortcomings and has even predicted the existence of new astrophysical objects, such as black holes \cite{2}. However, to match observations, the introduction of two (so far) elusive components—dark matter and dark energy—is necessary \cite{3}. Dark energy is essential for explaining the universe’s accelerated expansion, while dark matter is required to account for galactic rotation curves, gravitational lensing effects, and the large-scale structure of the cosmos. Despite their crucial role in ensuring accurate predictions, these dark components have not been directly detected to date, and their very nature remains one of the greatest mysteries of contemporary physics.

%Newtonian mechanics is foundational for understanding classical gravitational phenomena but encounters significant limitations in certain astrophysical contexts \cite{1}. General relativity has successfully addressed many of these limitations by providing a more comprehensive framework for understanding gravity \cite{2}. However, general relativity cannot fully explain some cosmic observations. To account for these, the concepts of dark matter and dark energy have been introduced \cite{3}. While dark matter remains undetected, its existence is inferred from gravitational effects on galaxies and galaxy clusters. These observations highlight the ongoing need for new theoretical frameworks in astrophysics.

Another way to reconcile theory with observations is to move beyond the standard gravity paradigm by exploring extensions of gravity. These alternative frameworks are constructed to reproduce the predictions of standard gravity where applicable while naturally generating effects that mimic dark matter and dark energy on cosmological scales\footnote{We note in passing that, beyond observational motivations, there are also compelling theoretical reasons to modify GR. One of the strongest is that the Hilbert-Einstein action, while the simplest formulation of gravity, is by no means the only possible one. This has led to various generalizations, particularly in the pursuit of unifying all fundamental interactions. Moreover, quantum corrections in the strong-field regime, introduced to ensure renormalizability, naturally extend the Einstein-Hilbert action by incorporating higher-order terms.}. Several alternatives for modifying gravity have been put forward. Some approaches involve generalizing the Einstein-Hilbert action \cite{p1}, while others introduce scalar fields \cite{p2,p3}, incorporate new geometric ingredients \cite{p4}, or treat physical constants as dynamical quantities \cite{p5,p6}. Among the simplest and most extensively studied modifications are $f(R)$ gravity theories, which replace the Ricci scalar in the action functional with a more general function of $R$ \cite{p7,p8}.

Besides these, other approaches have been developed specifically as extensions of Newtonian gravity. Perhaps the most well-known is Modified Newtonian Dynamics (MOND), which modifies Newtonian gravity when the acceleration of a test mass falls below a certain threshold \cite{Milgrom1}. 
Other proposals introduce various corrections to the standard $1/r$ gravitational potential, including power-law modifications, such as the model proposed by Maneff \cite{Maneff,Maneff2,Maneff3}, logarithmic terms, originally introduced by Tohline \cite{Tohline} and by Kuhn and Kruglyak \cite{Kuhn}, or exponential corrections, as seen in the Yukawa model \cite{Smullin} or certain forms of nonlocal gravity \cite{Nonlocal}. Such corrections can be motivated from different perspectives, including nonlocal extensions of the Poisson equation \cite{Blome,Blome2} and entropic gravity scenarios \cite{Ourabah}.

Identifying the correct theory among alternative gravity models requires comparing their phenomenological predictions with observational data. This allows for constraining their parameters and assessing their viability. In the context of modified Newtonian gravity or the weak-field limit of extensions to GR, the Jeans instability is a particularly relevant phenomenon that can serve as a test. It is responsible for the collapse of a gravitationally bound system, such as interstellar gas, ultimately leading to star formation. This mechanism has been extensively studied within the framework of various alternative gravity theories \cite{J1,J2,J3,J4,J5,J6,J7,J8}, where it often provides a valuable tool for constraining their parameters.

In this paper, we develop a general formulation of the Jeans instability within the framework of modified gravity. Our approach is sufficiently versatile to incorporate arbitrary corrections to the Newtonian potential, making it adaptable to a wide range of gravitational theories in the weak-field limit. The model is initially developed using a classical hydrodynamic approach, which is then extended to a quantum hydrodynamic framework. This provides insight into how thermal and quantum corrections influence and saturate the gravitational instability in the context of modified gravity.

We examine in detail three specific types of corrections to the Newtonian potential: Maneff corrections, which introduce an additional self-potential that scales as $1/r^2$; logarithmic corrections inspired by Tohline; and Yukawa-type corrections, characterized by exponential modifications. These three cases encompass a wide variety of behaviors that such corrections can exhibit. For each case, we derive criteria for the stability of a bound system within the framework of these modified gravity theories. We also constrain the parameters of these corrections by comparing the established stability criteria with the stability data of Bok globules—dense, cold regions of molecular clouds that are potential sites for star formation. These globules serve as an ideal testing ground due to their mass being on the order of their Jeans mass \cite{4}.

The paper is structured as follows: In Section \ref{Sec2}, we present a general formulation of the problem and derive the resulting dispersion relation, accounting for an arbitrary correction to the Newtonian potential. In Section \ref{Sec3}, we apply this approach to three specific corrections: Maneff, logarithmic, and Yukawa potentials. In Section \ref{Sec4}, we use data from Bok globules to constrain the parameters of these corrections and compare them with other independent bounds available in the literature. In Section \ref{Sec5}, we extend our approach to the quantum realm. Finally, in Section \ref{Sec6}, we conclude and suggest possible directions for future research.

\section{General formulation: dispersion relations in alternative gravity}\label{Sec2}

The Jeans instability is a gravitational instability that occurs in a self-gravitating gas cloud when the internal pressure is insufficient to counteract gravitational attraction. In the context of standard gravity, the mechanism can be formulated through the hydrodynamic equations coupled with the Poisson equation \cite{5,6}. Here, we consider a classical self-gravitating medium, where the self-gravitational potential is given by the Newtonian potential supplemented by an arbitrary correction term that depends only on the coordinate $r$. We restrict our analysis to a static universe, which is sufficient to derive the stability criteria. The extension to an expanding universe is conceptually straightforward and will be briefly discussed later. The dynamics of the medium is governed by the continuity equation, which ensures mass conservation,
 
\begin{equation}
\frac{\partial \rho}{\partial t} + \nabla \cdot (\rho \mathbf{u}) = 0,
\label{continuity}
\end{equation}  
 where \(\rho(\mathbf{r}, t)\) represents the mass density and \(\mathbf{u}(\mathbf{r}, t)\) is the velocity field, and the Euler (momentum-balance) equation,  
	\begin{equation}
			\frac{\partial \mathbf{u}}{\partial t} + (\mathbf{u} \cdot \nabla) \mathbf{u} = - \frac{\nabla}{m} \int \rho(\mathbf{r'}) \Phi(|\mathbf{r} - \mathbf{r'}|) \, d\mathbf{r'} - \frac{\nabla p}{\rho},
		\label{integrodifferential}
		\end{equation}
where $p$ is the pressure and $\Phi$ is an arbitrary self-potential. To keep the discussion as general as possible at this stage, we regard $\Phi$ as a (generic) self-potential, which can be written as

\begin{equation}
\Phi  \left(\mathbf{r}-\mathbf{r}^{\prime}\right) \equiv - \frac{ {G} m}{\left|\mathbf{r}-\mathbf{r}^{\prime}\right|} + \Phi_c  \left(\mathbf{r}-\mathbf{r}^{\prime}\right),
\label{pot}\end{equation}
where $\Phi_c$ is an arbitrary correction to the standard Newtonian self-potential. To close the system of equations (\ref{continuity})-(\ref{integrodifferential}), an equation of state is needed to determine the pressure 
$p$. Here, we consider a \textit{barotropic} fluid, where the pressure depends only on the density. That is,
\begin{equation}
p(\mathbf{r}, t) = p[\rho(\mathbf{r}, t)].
\label{pression}
\end{equation}

We restrict our analysis to the linear regime, where we consider small perturbations superimposed on a stationary, infinite, homogeneous, and isotropic equilibrium medium. By assuming that these perturbations remain sufficiently small, we can linearize the governing equations, allowing us to study the system's response without invoking nonlinear effects. This approach enables us to derive a dispersion relation that characterizes the stability of the system under different modifications to the gravitational potential. We thus write
		\begin{equation}
			\begin{aligned}
			    \rho(\mathbf{r}, t) &= \rho_0 + \delta\rho(\mathbf{r}, t), \\
			    \mathbf{u}(\mathbf{r}, t) &= \mathbf{u}_0 + \delta\mathbf{u}(\mathbf{r}, t).
		\end{aligned}
		\end{equation}
		Here, $\rho_0$ and $\mathbf{u}_0$ represent the equilibrium quantities, while $\delta\rho(\mathbf{r}, t)$ and $\delta\mathbf{u}(\mathbf{r}, t)$ correspond to the small perturbations. After linearization of equations \eqref{continuity} and \eqref{integrodifferential}, we obtain\footnote{Here, we have implicitly assumed, as is customary, that the self-potential is sourced only by the perturbation $\delta \rho$ and not by the background density $\rho_0$, a procedure known as the \textit{Jeans swindle}. While this procedure may seem \textit{ad hoc}, it has been shown to be mathematically sound (see \cite{H1,H2}). Alternatives, to avoid the Jeans swindle, include introducing a background potential, considering cosmic expansion, or finite-domain effects \cite{H3,H4}.}
		\begin{equation}
			\begin{aligned}
				\frac{\partial \delta \rho(\mathbf{r}, t)}{\partial t} &+ \rho_0 \nabla \cdot \delta \mathbf{u}(\mathbf{r}, t) = 0, \\
				\frac{\partial \delta \mathbf{u}(\mathbf{r}, t)}{\partial t} &= - \frac{\nabla}{m} \int \delta \rho(\mathbf{r'}, t) \Phi(|\mathbf{r} - \mathbf{r'}|) \, d\mathbf{r'} - c_s^2 \frac{\nabla \delta \rho(\mathbf{r}, t)}{\rho_0},
			\label{perturbations}
			\end{aligned}
		\end{equation}
where 

		\begin{equation}
			c_s^2 \equiv \left( \frac{dp}{d\rho} \right)_{\rho = \rho_0}
			\label{speed}
		\end{equation}
		is the squared speed of sound in the medium. By performing a Fourier transform on the equations \eqref{perturbations} and combining the resulting expressions, we obtain the dispersion relation
		
		\begin{equation}
		    \omega^2 = \frac{\rho_0}{m} \tilde{\Phi}(k) k^2 + c_s^2 k^2,
		    \label{dispersion}
		\end{equation}
		where
		
		\begin{equation}
		    \tilde{\Phi}(k) = \int \Phi(r) \exp(-i \mathbf{k} \cdot \mathbf{r}) \, d\mathbf{r}
		    \label{fourier}
		\end{equation}
is the Fourier transform of the self-potential $\Phi$. For a general self-potential given by Eq. (\ref{pot}), by explicitly expressing the Fourier transform of the Newtonian component of the potential, we have 
\begin{equation}
\tilde{\Phi}(k) = -  \frac{4 \pi G m }{k^{2}}+ \widetilde{\Phi_c},
\label{fourier}
\end{equation}
where     
\begin{equation}
\widetilde{\Phi_c} \equiv  \int \Phi_c(r)  \exp(-i \mathbf{k} \cdot \mathbf{r}) \, d\mathbf{r}.
\end{equation}
The generic dispersion relation (\ref{dispersion}) rewrites as\footnote{Throughout this paper, we focus on the case of a static universe, which is sufficient for deriving the gravitational stability criterion. Nonetheless, the extension to an expanding cosmological background is straightforward (see, e.g., \cite{Bonnor}). In such a scenario, the evolution of the density contrast $\delta(t) := \delta \rho(t)/\rho_0$ is governed by
\begin{equation}\label{Bc} \ddot{\delta} + 2 \frac{\dot{a}}{a} \dot{\delta} + \left[ - \Omega_J^2 + \frac{\rho_0}{m} \widetilde{\Phi_c} \left(\frac{k}{a} \right) \frac{k^2}{a^2} + \frac{c_s^2 k^2}{a^2}   \right] \delta = 0, \end{equation}
where $a(t)$ denotes the cosmological scale factor. This equation generalizes the classical Bonnor equation \cite{Bonnor} to incorporate corrections arising from modified gravitational interactions. In the static limit ($a = 1$), one recovers the dispersion relation (\ref{dispersion final}) by assuming a harmonic time dependence of the form $\delta(t) \propto \exp(-i \omega t)$.
}
		
\begin{equation}
\omega^2 = -\Omega_J^2 + \frac{\rho_0}{m} \widetilde{\Phi_c}(k) k^2 + c_s^2 k^2,
\label{dispersion final}
\end{equation}
where $\Omega_J \equiv \sqrt{4\pi G \rho_0}$  is the so-called Jeans frequency. In the absence of corrections ($\Phi_c = 0$), Equation (\ref{dispersion final}) correctly recovers the standard gravitational dispersion relation, as expected, describing the balance between Newtonian attraction and the opposing pressure forces that stabilize the system.

Instability is characterized by a nonzero imaginary part of the frequency. To make this explicit, we write $ \omega $ as $ \omega \equiv \omega_r + i \omega_i $. Since $ \omega^2 $ is real, the frequency is either purely real or purely imaginary. The transition between these two cases occurs when $ \omega = 0 $, which defines a critical wave number. In the standard Newtonian case, the Jeans wave-number $k_J$ reads as
\begin{equation}
k_J=\frac{\Omega_J}{c_s}.
\end{equation}
In the more general scenario, depending on the sign of $\Phi_c$ (whether it introduces an additional attractive or repulsive contribution) and its functional dependence on the coordinate $r$, the correction may either enhance the instability or, conversely, help suppress it. In the next Section, we will examine three specific examples of potential corrections $\Phi_c$, each corresponding to a different alternative theory of gravity. These cases encompass three general types of modifications that are likely to occur: power-law, logarithmic, and exponential corrections. 

\section{Specific Examples}\label{Sec3}

		In this section, we explore in detail specific corrections to the gravitational potential that have been proposed in the literature to resolve observational discrepancies or improve theoretical models in various contexts. We focus on the Maneff, logarithmic, and Yukawa potentials, each providing distinct modifications to Newtonian gravity, and derive their corresponding dispersion relations and Jeans wave-numbers.
		
		\subsection{Maneff Potential}

		The first case we discuss is the Maneff (sometimes spelled Manev) potential, introduced as a classical alternative to general relativity \cite{7}. This potential modifies the Newtonian gravitational potential by incorporating a correction term that scales as $1/r^2$, offering a way to account for relativistic-like effects without the need for the full formalism of Einstein's theory \cite{8,9,10}. This potential has recently been found to arise in a Newtonian-like theory inspired by the Brans–Dicke gravitational Lagrangian \cite{Man}. It is expressed as\footnote{Note that this scenario can be regarded as a specific instance of modified gravity models incorporating power-law corrections of the form $\propto 1/r^{\alpha}$ to the Newtonian potential—modifications that may naturally arise, for example, within the framework of entropic gravity \cite{Sheykhi}.}

	\begin{equation}
		\Phi = -\frac{Gm}{r} \left( 1 + \frac{B}{r} \right),
		 \label{maneff}
	\end{equation}
where $B$ is a correction parameter with the dimension of length. In the original Maneff proposal, the parameter $B$ is given by $B = {3Gm}/{c^2}$. In this case, the Fourier transform of the correction potential $\Phi_c$ reads as

\begin{equation}
\widetilde{\Phi_c}(k) = - Gm B\int \frac{\exp(-i \mathbf{k} \cdot \mathbf{r})}{r^2}  \, d\mathbf{r} = -  \frac{2\pi^2 Gm B}{k}.
\label{fourier maneff}
\end{equation}

The generic dispersion relation \eqref{dispersion final} reduces accordingly to

\begin{equation}
\omega^2 = -\Omega_J^2 \left(1+  \frac{\pi }{2} B k \right) + c_s^2 k^2.
\label{dispersion maneff}
\end{equation}

The threshold value of $k$, which separates the oscillatory regime from the instability regime, is obtained by setting $\omega = 0$ and is given by

\begin{equation}
\tilde{k_J} = \frac{\pi \Omega_J^2 B + \Omega_J \sqrt{16 c_s^2 + \pi^2 \Omega_J^2 B^2}}{4 c_s^2}.
\label{k maneff}
\end{equation}
Perturbations with wavenumbers smaller than $\tilde{k_J}$ lead to instability, while those with wavenumbers greater than 
$\tilde{k_J}$ result in oscillatory behavior. Normalizing to the standard Jeans wavenumber $k_J$, one has

\begin{equation}
\frac{\tilde{k_J}}{k_J} = \frac{\pi B}{4} k_J + \sqrt{1 + \frac{\pi^2 B^2}{16}k_J^2}.
\end{equation}

\begin{figure}
    \centering
    \includegraphics[width=0.5\linewidth]{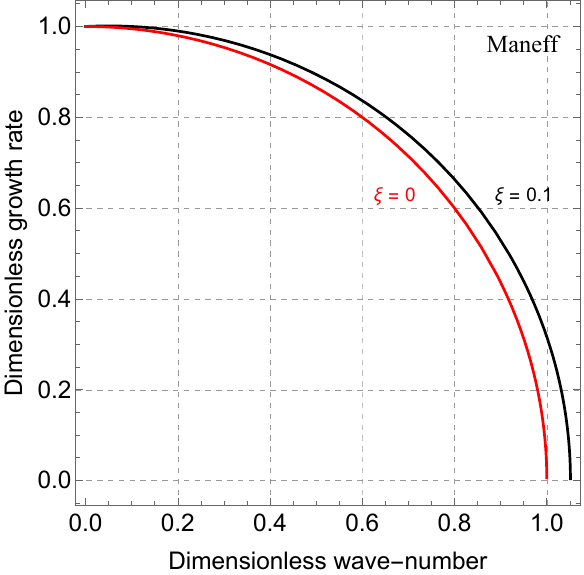}
    \caption{Dimensionless growth rate for Maneff potential with $\xi=0.1$ compared to the Newtonian growth rate ($\xi=0$).}
    \label{figManeff}
\end{figure}

For $B > 0$, this ratio exceeds unity, indicating that instability sets in for perturbations with shorter wavelengths compared to the Newtonian case. The dispersion relation (\ref{dispersion maneff}) can be rewritten in a dimensionless form as

\begin{equation}
W^2= -1 -\xi K+ K^2
\end{equation}
where we have defined
\begin{equation}
    W:=\omega / \Omega_J, \quad K:=c_s k/\Omega_J, \quad \xi:= \pi \Omega_J B/2 c_s.
\end{equation}
The growth rate, namely the imaginary part of $\omega$, reads in dimensionless form, i.e. $\Gamma:= {Im}(\omega)/\Omega_J$ as follows

\begin{equation}\label{G1}
 \Gamma= \sqrt{1+\xi K-K^2 }. 
\end{equation}
The latter is depicted in Fig. \ref{figManeff} for different values of $\xi$. It can be seen that the growth rate is higher in this scenario compared to Newtonian gravity, while the maximum growth rate, $\sqrt{4\pi G \rho_0}$, remains the same for both potentials and is attained in the infinite wavelength limit ($k=0$).

	\subsection{Logarithmic Potential}

	The second type of modified gravitational potential we consider includes a logarithmic correction. This form was first introduced by Tohline \cite{Tohline}, who showed that such a correction could stabilize a cold stellar disk. Kuhn and Kruglyak, along with other authors \cite{12,13,14,Kuhn,16}, further developed this phenomenological approach, emphasizing its applicability in various astrophysical contexts. This potential reads as
		\begin{equation}
			\Phi=-\frac{Gm}{r}-\delta Gm\ln{\left(\eta r\right)},
			\label{log}
		\end{equation}
where $ \delta $ is a parameter with dimension inverse to length and $ \eta $ is an arbitrary constant. In this case, the Fourier transform of the correction reads as\footnote{Note that the Fourier transform of the logarithmic potential is defined in the sense of distributions using the Cauchy principal value (see e.g. \cite{m,17} for details).}  

		\begin{equation}
			\widetilde{\Phi_c}(k) = -Gm\delta \int  \ln \left(\eta r \right) \exp(-i \mathbf{k} \cdot \mathbf{r}) \, d\mathbf{r} =   \frac{2\pi^2 Gm  \delta}{k^3}.
			\label{fourier log}
		\end{equation}

Thus, the dispersion relation (\ref{dispersion final}) becomes 

\begin{equation}
\omega^2 = - \Omega_J^2 \left( 1-\frac{\pi}{2} \frac{\delta}{k}\right)  + c_s^2 k^2.
\label{dispersion log}
\end{equation}

Analyzing the dispersion relation, one finds that $\delta$ must necessarily be negative to ensure that the corrective term effectively contributes to an attractive potential, in accordance with the theoretical expectations.

By setting $\omega = 0$ and solving for $k$, we obtain the critical wavenumber

		\begin{equation}
			\tilde{k_J} = \frac{2 \cdot 6^{1/3} \Omega_J^2 c_s^2 + \Delta^{2/3}}{6^{2/3} c_s^2 \Delta^{1/3}},
			\label{k log}
		\end{equation}

		where

		\begin{equation}
			\Delta := -9\pi \Omega_J^2 c_s^4 \delta + \sqrt{-48 \Omega_J^6 c_s^6 + 81\pi^2 \Omega_J^4 c_s^8 \delta^2}.
		\end{equation}
        
        To ensure a physical solution, we require the condition :

		\begin{equation}
            \delta \leq -\frac{4 \Omega_J}{3 \sqrt{3} \pi c_s}
			\label{delta condition}
		\end{equation}

        The ratio of the critical wavenumber $\tilde{k_J}$ to the standard Jeans wavenumber ${k_J}$ can be expressed as

		\begin{equation}
			\frac{\tilde{k_J}}{k_J} = \frac{2 \Omega_J c_s}{6^{1/3} \Delta^{1/3}} + \frac{\Delta^{1/3}}{6^{2/3} \Omega_J c_s}.
		\end{equation}

        It is easily verified that $\tilde{k_J}/k_J > 1$ when $\delta < 0$, indicating that gravitational instability develops at shorter wavelengths compared to the standard Newtonian case.

\begin{figure}
    \centering
    \includegraphics[width=0.5\linewidth]{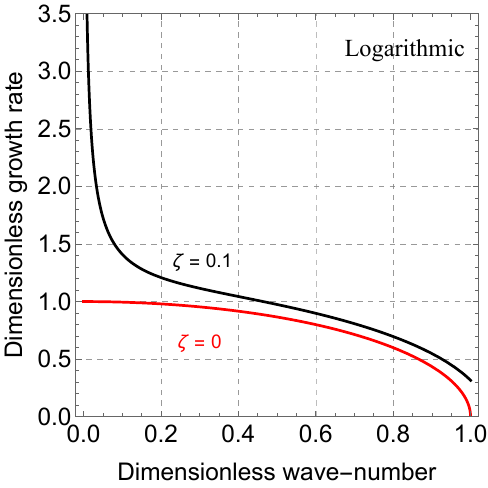}
    \caption{Dimensionless growth rate for logarithmic potential with $\zeta=0.1$ compared to the Newtonian growth rate ($\zeta=0$).}
    \label{figLog}
\end{figure}

Rewriting the dispersion relation (\ref{dispersion log}) in a dimensionless form, one has

\begin{equation}
W^2= -1-\frac{\zeta}{K}+ K^2,
\end{equation}
where we have defined 

\begin{equation}
    W:=\omega / \Omega_J, \quad K:=c_s k/\Omega_J, \quad \zeta:= - \pi c_s \delta /2 \Omega_J,
\end{equation}
The corresponding growth rate is shown in Fig. \ref{figLog}. We observe that, for the logarithmic potential, the growth rate diverges in the limit $k \to 0$ (infinite wavelengths). Moreover, for finite wavelengths ($k \neq 0$), it remains higher than in the Newtonian case.

        \subsection{Yukawa Potential}
The final example we consider is the Yukawa-type self-potential. Originally introduced by Yukawa in 1935 to describe the strong nuclear force between nucleons \cite{18}, this potential features an exponential decay with distance, rendering the force short-ranged rather than long-ranged. In the context of gravity, the Yukawa potential has been extensively studied, both as a small-scale correction to the Newtonian force \cite{19,20} and as a large-scale modification in alternative theories of gravity aimed at explaining astrophysical phenomena \cite{21,22,23}. The potential is given by
		\begin{equation}
			\Phi=-\frac{Gm}{r}\left(1+\alpha e^{-r / \lambda}\right)
			\label{yukawa}
		\end{equation}
		where $\alpha$ and $\lambda$ represent the strength and range of the correction, respectively. In this case, the Fourier transform of the correction reads as 

		\begin{equation}
			\widetilde{\Phi_c}(k) = - Gm\alpha \int  \frac{e^{-r/\lambda}}{r} \exp(-i \mathbf{k} \cdot \mathbf{r}) \, d\mathbf{r} = -  \frac{4\pi Gm \alpha \lambda^2}{1 + \lambda^2 k^2}.
            \label{fourrier yukawa}
		\end{equation}

		The dispersion relation (\ref{dispersion final}) becomes

		\begin{equation}
			\omega^2 = -\Omega_J^2 \left(1+\frac{\alpha \lambda^2 k^2}{1 + \lambda^2 k^2}\right)  + c_s^2 k^2.
            \label{dispersion yukawa}
		\end{equation}

	Setting $\omega = 0$ and solving the dispersion relation (\ref{dispersion yukawa}) yields the critical wavenumber

		\begin{equation}
			\tilde{k_J}^2 = \frac{-\chi + \sqrt{4 c_s^2 \lambda^2 \Omega_J^2 + \chi^2}}{2 c_s^2 \lambda^2},
            \label{k yukawa}
		\end{equation}

		where

		\begin{equation}
			\chi := c_s^2 - (1 + \alpha) \lambda^2 \Omega_J^2.
		\end{equation}

        To assess the impact of Yukawa corrections, we compare the modified critical wavenumber (\ref{k yukawa}) with the standard Jeans wavenumber. The ratio reads as

		\begin{equation}
			\frac{\tilde{k_J}^2}{k^2_J} = \frac{-c_s^2}{2 \lambda^2 \Omega_J^2} + \frac{1+\alpha}{2} + \sqrt{\frac{c_s^2}{\lambda^2 \Omega_J^2}+\left(\frac{c_s^2}{2 \lambda^2 \Omega_J^2}-\frac{1+\alpha}{2}\right)^2 }.
		\end{equation}  
		
		It is straightforward to verify that for $\alpha > 0$, the ratio $\tilde{k}_J / k_J$ exceeds unity, indicating that Yukawa corrections lead to instability at shorter wavelengths compared to the standard Newtonian case.

\begin{figure}
            \centering
            \includegraphics[width=0.5\linewidth]{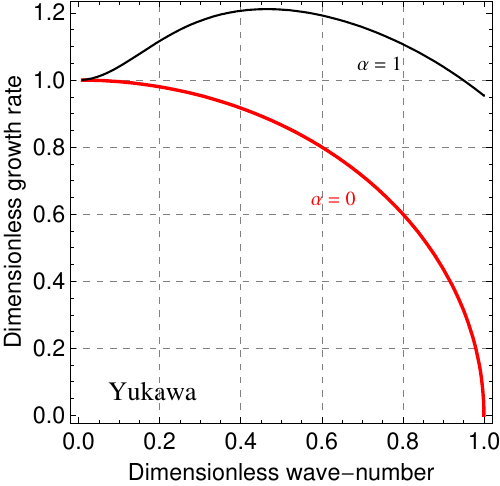}
            \caption{Dimensionless growth rate for the Yukawa potential  with $\alpha=1$ compared to the Newtonian growth rate ($\alpha=0$). For illustrative purposes, we have set $\gamma=0.1$.}
            \label{figYukawa}
        \end{figure}

        Rewriting the dispersion relation (\ref{dispersion yukawa}) in a dimensionless form, one has

        \begin{equation}
            W^2= -1-\frac{\alpha}{\frac{\gamma}{K^2}+1}+ K^2,
        \end{equation}
      where we have defined 
        
        \begin{equation}
            W:=\omega / \Omega_J, \quad K:=c_s k/\Omega_J, \quad \gamma:= c_s^2 / \Omega_J^2 \lambda^2.
        \end{equation}
      The corresponding growth rate is shown in Fig. \ref{figYukawa}. One may see that the growth rate is higher than in the Newtonian case, while the maximum value, $\sqrt{4\pi G \rho_0}$, remains unchanged and is attained at $k = 0$ (infinite wavelength).\\

Before closing this section, it is worth noting that the results obtained for all three potentials exhibit the same qualitative behavior: each correction leads to an increase in the Jeans wavenumber relative to the standard Newtonian case. This implies that the modified potentials enhance the instability of the system, lowering the threshold for gravitational collapse. As a result, structures can form at smaller wavelengths and with lower masses than those predicted by the classical Jeans criterion. This shared feature among the different corrections may help address the observed discrepancies in the stability of Bok globules, where observations appear inconsistent with predictions based on the standard Newtonian framework. These implications, along with constraints on the correction parameters, will be discussed in the next section.

\section{Data Analysis}\label{Sec4}

	In this section, we put our findings to the test by comparing them with available observational data on the stability of Bok globules. These small, dense, and isolated molecular clouds are typically sites of early star formation, making them excellent natural laboratories for probing alternative theories of gravity. Their masses are generally close to the classical Jeans mass, meaning that even slight deviations from Newtonian gravity can lead to significantly different predictions regarding their stability.

{To better account for 
the observational data reported in \cite{24,25}}, we apply the modified stability criteria derived in the previous section, aiming to adjust the predicted Jeans mass accordingly. This comparison allows us to place constraints on the parameters characterizing the modified gravitational potentials.

Table \ref{observation} presents data for 11 Bok globules, compiled from \cite{24} and also discussed in \cite{25}. Notably, in 7 of these cases, there is a clear mismatch between observational evidence and theoretical expectations: the globules are undergoing star formation—implying gravitational instability—despite having masses below their classical Jeans mass. Several proposals have been put forward in the literature to account for this discrepancy, including the influence of dark matter \cite{Our1} and modifications to the underlying gravitational theory \cite{25,J5}, and other alternative mechanisms \cite{M1,M2}. In what follows, we examine to what extent the modified potentials discussed above can provide a plausible explanation.

		\begin{table}[h]
		    \centering
		   \setlength{\tabcolsep}{10pt} 
		    \begin{tabular}{lcccccc}
            \hline \hline
		        
		        Name & $T$ (K) & $n_{\text{H}_2}$ (cm$^{-3}$) & $M$ (M$_\odot$) & $M_J$ (M$_\odot$) & Stability \\
		        \hline
		        CB 87   & 11.4 & $(1.7 \pm 0.2) \times 10^4$ & $2.73 \pm 0.24$ & 9.6  & Stable     \\
		        CB 110  & 21.8 & $(1.5 \pm 0.6) \times 10^5$ & $7.21 \pm 1.64$ & 8.5  & Unstable   \\
		        CB 131  & 25.1 & $(2.5 \pm 1.3) \times 10^5$ & $7.83 \pm 2.35$ & 8.1  & Unstable   \\
		        CB 134  & 13.2 & $(7.5 \pm 3.3) \times 10^5$ & $1.91 \pm 0.52$ & 1.8  & Unstable   \\
		        CB 161  & 12.5 & $(7.0 \pm 1.6) \times 10^4$ & $2.79 \pm 0.72$ & 5.4  & Unstable   \\
		        CB 184  & 15.5 & $(3.0 \pm 0.4) \times 10^4$ & $4.70 \pm 1.76$ & 11.4 & Unstable   \\
		        CB 188  & 19.0 & $(1.2 \pm 0.2) \times 10^5$ & $7.19 \pm 2.28$ & 7.7  & Unstable   \\
		        FeSt 1-457 & 10.9 & $(6.5 \pm 1.7) \times 10^5$ & $1.12 \pm 0.23$ & 1.4 & Unstable   \\
		        Lynds 495  & 12.6 & $(4.8 \pm 1.4) \times 10^4$ & $2.95 \pm 0.77$ & 6.6  & Unstable     \\
		        Lynds 498  & 11.0 & $(4.3 \pm 0.5) \times 10^4$ & $1.42 \pm 0.16$ & 5.7  & Stable     \\
		        Coalsack   & 15.0 & $(5.4 \pm 1.4) \times 10^4$ & $4.50$ & 8.1  & Stable     \\
		        \hline \hline
		    \end{tabular}
             \caption{Selected Bok globules, their kinetic temperatures, particle number densities, masses, conventional Jeans masses, and observed stability, as reported in \cite{24}.}
		    \label{observation}
		\end{table}
		
		Based on the critical wavenumbers derived in the previous section, it is more appropriate to evaluate the corresponding critical mass—namely, the modified Jeans mass. This is defined as the mass contained within a sphere of diameter $2\pi / \tilde{k_J}$, that is,
\begin{equation}
\tilde{M_J} = \frac{4\pi \rho_0}{3} \left(\frac{\tilde{\lambda_J}}{2}\right)^3,
\label{jeans mass}
\end{equation}
where $\tilde{\lambda_J} = {2\pi}/{\tilde{k_J}}$ is the modified Jeans wavelength. Since the Jeans mass scales as $k_J^{-3}$, we can establish the relationship between the modified Jeans mass $\tilde{M_J}$ and a reference Jeans mass  $M_J$ as

		\begin{equation}
			\tilde{M_J} = \left(\frac{k_J}{\tilde{k_J}}\right)^3 M_J.
			\label{mass ratio}
		\end{equation}

To derive constraints on the parameters of the generalized gravitational potentials, we use the data from Table \ref{observation}. Specifically, we focus on 7 Bok globules for which the observed behavior contradicts the predictions based on the standard Jeans criterion. These globules exhibit gravitational instability (evidenced by ongoing star formation) despite having masses smaller than their corresponding Jeans mass. This apparent discrepancy suggests that the classical Jeans mass is insufficient to describe the stability of these systems in the presence of modified gravitational theories. To reconcile the theoretical predictions with the observed data, we impose the condition that the modified Jeans mass must be at least equal to the observed mass of each globule. This constraint ensures that the system remains gravitationally unstable under the modified potential. By applying this requirement, we can derive bounds on the ratio of the modified Jeans wavenumber to the standard Jeans wavenumber, $(k_J/\tilde{k_J})^3$. These bounds provide a direct way to constrain the parameters of the modified gravitational potentials, offering insights into how deviations from Newtonian gravity may influence the stability of self-gravitating systems such as Bok globules.

On the other hand, for Bok globules whose observed stability aligns with the predictions of the standard Jeans criterion, any modification to the gravitational potential must preserve this agreement. In other words, the modified potential should not alter the expected stability in these cases. This requirement imposes additional constraints on the parameters of the modified potentials.

%		To match the observations of Table \ref{observation}, we need to adjust the ratio $\left(\frac{k_J}{\tilde{k_J}}\right)^3$. To achieve this, we will study the ratio $\frac{M}{M_J}$, which provides an upper or lower bound that the ratio $\left(\frac{k_J}{\tilde{k_J}}\right)^3$ must respect.

%		For a stable globule, the ratio $\frac{M}{M_J}$ represents the lower bound that $\left(\frac{k_J}{\tilde{k_J}}\right)^3$ must not exceed. On the other hand, for an unstable globule, the ratio $\frac{M}{M_J}$ represents the upper bound that $\left(\frac{k_J}{\tilde{k_J}}\right)^3$ must not exceed.
%		For the Maneff potential, we have the following ratio, which provides a constraint on $B$:
%		\begin{equation}
%		    \left(\frac{k_J}{\tilde{k_J}}\right)^3 = \left(\frac{4c_s}{\pi \Omega_J B + \sqrt{16 c_s^2 + \pi^2 \Omega_J^2 B^2}}\right)^3
%			\label{k maneff ratio}
%		\end{equation}

%		For the logarithmic potential, we have condition \eqref{delta condition}, along with the following ratio to impose a constraint on $\delta$ :
%		\begin{equation}
%		    \left(\frac{k_J}{\tilde{k_J}}\right)^3 = \left(\frac{6^{2/3} c_s \Delta^{1/3} \Omega_J}{2 \cdot 6^{1/3} \Omega_J^2 c_s^2 + \Delta^{2/3}}\right)^3
%			\label{k log ratio}
%		\end{equation}
				
%		Finally, for the Yukawa potential, the equation is:
		
%		\begin{equation}
%		    \left(\frac{k_J}{\tilde{k_J}}\right)^3 = \left(\frac{2 \Omega_J^2 \lambda^2}{-\xi + \sqrt{4 c_s^2 \lambda^2 \Omega_J^2 + \xi^2}}\right)^{3/2}
%		\end{equation}

The results are summarized in Table \ref{results}, which lists the acceptable parameter values required to match the observational data. Specifically, the table provides the allowed range for the parameter $B$ in the case of the Maneff potential and for $\delta$ in the case of logarithmic corrections. In the case of the Yukawa potential, which involves two free parameters, the data in Table \ref{results} are obtained by fixing the strength parameter to $\alpha = 1$, as is commonly done in experimental tests \cite{20}.  More generally, however, agreement with the observational data can be achieved through an appropriate balance between the two Yukawa parameters, $\alpha$ and $\lambda$. This interplay is illustrated in Figure \ref{yukawa} for the specific case of the Bok globule CB 184.
        
	%	This expression depends on two parameters, $\lambda$ and $\alpha$ (included in $\xi$), so the resulting ratio is represented in 3D, as shown in Figure \ref{yukawa}. Different pairs of $\lambda$ and $\alpha$ values can yield good results. However, as observed in Figure \ref{yukawa}, values of $\alpha$ below 0.8 cannot be adjusted to provide results that match the observations. In this study, we will focus on the value $\alpha = 1$ and determine the corresponding value of $\lambda$ that fits the observations.

\begin{figure}[h]
\centering
\includegraphics[width=1\linewidth]{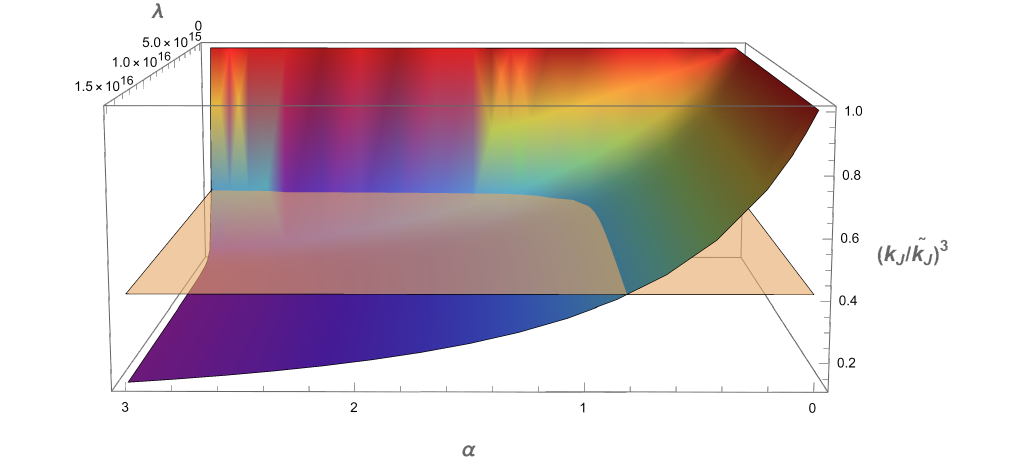}
\caption{The ratio $\left({k_J}/{\tilde{k}_J}\right)^3$ as a function of the Yukawa parameters $\alpha$ and $\lambda$ for the Bok globule CB 184. The transparent plane indicates the threshold value of the ratio required for the modified Jeans mass to be consistent with observational data. Parameter combinations that yield a ratio below this plane correspond to scenarios in which the modified gravitational potential accounts for the observed instability of the globule.}
\label{yukawa}
\end{figure}

\begin{table}[h]
\centering
\setlength{\tabcolsep}{10pt} 
\begin{tabular}{lcc|c|c|cc}
\hline \hline
Bok Globule & $ {M}/{M_J} $ & Stability & $B$ ($10^{13}$ m) & $\delta$ ($10^{-16}$m$^{-1}$) & $\lambda$ ($10^{14}$ m) $\left(\alpha=1\right)$ \\
\hline
CB 87  & 0.284  & Stable & $\leq 71.0 $  &  $-9.69\leq  \delta \leq -1.88 $  & $\mathbb{R}$ \\
CB 110  & 0.848  & Unstable & $\geq 4.30$ & $\leq -4.04 $& $\geq 2.10$ \\
CB 131  & 0.966  & Unstable & $\geq 0.86$ & $\leq -4.86 $ & $\geq 0.83 $\\
CB 134  & 1.06   & Unstable  & $\mathbb{R}^+$ & $\leq -11.7 $& $\mathbb{R}$ \\
CB 161  & 0.516  & Unstable & $\geq 19.1 $& $\leq -6.56 $ & $\geq 6.03$ \\
CB 184  & 0.412  & Unstable & $\geq 43.8 $ & $\leq -6.03 $  & $\geq 17.5 $  \\
CB 188  & 0.933  & Unstable & $\geq 2.01 $ &  $\leq -3.87 $ & $\geq 1.44$\\
FeSt 1-457  & 0.8   & Unstable & $\geq 2.24 $ & $\leq -11.9$ & $\geq 0.91$\\
Lynds 495  & 0.446  & Unstable & $\geq 28.3 $ & $\leq -7.31 $& $\geq 9.90$ \\
Lynds 498  & 0.249  & Stable & $\leq 53.6$ &  $-19.0 \leq \delta \leq -3.05 $& $\mathbb{R}$\\
Coalsack  & 0.555  & Stable & $\leq 21.0$ &  $-4.42 \leq \delta \leq -2.93 $& $\leq 6.60$ \\
\hline \hline
\end{tabular}
\vspace{0.5em}
\caption{The ratio $M/M_J$ and the corresponding ranges of the parameters $B$ (for the Maneff potential), $\delta$ (for the logarithmic correction), and $\lambda$ (for the Yukawa potential with fixed $\alpha = 1$) required to match the observational data presented in Table \ref{observation}.}
\label{results}	
\end{table}

Before concluding this section, it is useful to compare the parameter constraints derived in this study with those reported in the existing literature. In its original form, the Maneff potential assumes $B ={3GM}/{c^2}$, a value that cannot readily achieve the constraints required by our analysis for Bok globules. However, this definition of $B$ can be relaxed if the potential is considered from alternative theoretical perspectives \cite{Sheykhi,Man}. Regarding the logarithmic potential, there is no widely accepted value for the parameter $\delta$. The values reported in the literature vary significantly, differing by several orders of magnitude between studies (see Table \ref{T3}). Finally, for the Yukawa potential, the literature presents a broad range of parameter values. When interpreted as a short-range modification to Newtonian gravity, experimental constraints limit $\lambda < 56,\mu\text{m}$ (imposing $|\alpha| = 1$) \cite{20}. In contrast, when viewed as a long-range interaction \cite{26}, typical values involve a much larger range, $\lambda \sim 10^{15}$ m, similar to those found in our analysis, but associated with much weaker strengths, typically $\alpha \sim 10^{-9}$. The wide variation in parameters across different physical scenarios indicates that these modified potentials are not universally applicable to all scales and should not be considered as general substitutes for Newtonian gravity. Instead, they should be regarded as effective models designed to capture specific gravitational behaviors in particular physical contexts.

			\begin{table}[h!]
			    \centering
			    \caption{\textbf{Summary of Limits on $\delta$ as reported in \cite{16}}}
			    \label{limits_alpha}
			    \begin{tabular}{l c  l}
			        \hline \hline
			        & $\delta$ ($10^{-21}$ m$^{-1})$ & \textbf{Adopted data} \\
			        \hline
			        Chang \& al. & $(6.16 \pm 4.54) $  & Five binary pulsars \cite{16}\\
			        Fabris \& Campos (2009) & $\sim -3.24$  & Rotation curves of 10 spiral galaxies \cite{12} \\
			        Deng \& Xie (2014) & $\sim -3.24 \times 10^{-3}$  & Planetary motions in the solar system \cite{Deng} \\
			        \hline \hline
			    \end{tabular}
                \label{T3}
			\end{table}

\section{Quantum Regime}\label{Sec5}

        Up to this point, we have investigated gravitational instabilities in a classical framework, where pressure acts as the primary force opposing gravitational collapse. However, in astrophysical settings involving very low temperatures, high densities, or small mass scales—such as in the early universe, compact astrophysical objects, or certain dark matter models—quantum effects can become significant and modify the classical stability criteria. It is therefore instructive to extend our analysis to the quantum regime, where quantum pressure and wave-like behavior of matter may influence the onset and evolution of gravitational instabilities. This will shed light on the behavior of quantum matter confined by gravitational forces in the context of alternative gravitational models.
        
      The standard framework for exploring quantum gravitational instability relies on the Schrödinger–Newton equation (also referred to as the Schrödinger–Poisson equation). This equation has a rich history, originally introduced by Ruffini and Bonazzola in the context of self-gravitating boson stars, where it provides a semi-classical bridge between quantum mechanics and gravity \cite{27}. Later, Diósi and Penrose \cite{28,29} proposed it as a phenomenological model for wavefunction collapse, highlighting its potential relevance for probing the interface between quantum theory and gravitation. The Schrödinger–Newton equation couples the linear Schrödinger equation, which governs the evolution of the wave function $\psi$, with the nonlinear Poisson equation for the gravitational potential generated by the mass density associated with $\psi$ \cite{30}. That is,

		\begin{equation}
			i\hbar \frac{\partial \psi}{\partial t} = \left(-\frac{\hbar^2}{2m} \nabla^2 + m\Phi \right) \psi,
			\label{schrodinger}
		\end{equation}
		\begin{equation}
			\nabla^2 \Phi = 4\pi G m |\psi|^2,
			\label{poisson SN}
		\end{equation}

which can be rewritten as a single integro-differential equation, as follows\footnote{We note in passing that, given the implications of the original Schrödinger-Newton equation in addressing conceptual issues, such as the wavefunction collapse, such generalizations, enhanced with a corrective potential (\ref{pot}), could have interesting consequences at a fundamental level, warranting independent consideration.}

		\begin{equation}
			i\hbar \frac{\partial \psi}{\partial t} = \left( -\frac{\hbar^2}{2m} \nabla^2 - m \int \Phi(|\mathbf{r} - \mathbf{r}'|) |\psi(\mathbf{r}', t)|^2 \, d^3r' \right) \psi.
			\label{SN equation}
		\end{equation}
As in the classical analysis, we relax the assumption that the self-potential is strictly Newtonian and instead consider a generic form that incorporates corrections to the standard Newtonian potential. That is, $\Phi$ takes the form given in Eq. (\ref{pot}).

To investigate the Jeans gravitational instability in this framework, it is convenient to reformulate the Schrödinger–Newton equation in hydrodynamic form via the Madelung transformation\footnote{See Appendix \ref{A} for an alternative kinetic treatment using the Wigner function.} \cite{32}. This is achieved by expressing the wave function $\psi$ in polar form

	\begin{equation}
			\psi(r, t) = A(\mathbf{r}, t)e^{i S(\mathbf{r},t)/\hbar},
			\label{madelung}
		\end{equation}
where $A(\mathbf{r}, t)$ represents the amplitude and $S(\mathbf{r}, t) = \frac{\hbar}{2i} \ln\left({\psi}/{\psi^*}\right)$ denotes the phase of the wave function. Since we are dealing with a purely gravitational system, it is natural to normalize the wave function to the mass density. That is, 
\begin{equation}
\rho(\mathbf{r}, t) \equiv |\psi|^2 = A(\mathbf{r}, t)^2,	
\label{density}
\end{equation}
and the velocity field is defined by\footnote{Note that, as defined, the velocity field is irrotational, i.e., $\nabla \times \mathbf{u} = 0$.}

		\begin{equation}
			\mathbf{u} \equiv \frac{\nabla S}{m} = \frac{i \hbar}{2m} \frac{\psi \nabla \psi^* - \psi^* \nabla \psi}{|\psi|^2}.	
			\label{velocity}
		\end{equation}
By substituting the wave function \eqref{madelung} into the Schrödinger-Newton equation \eqref{SN equation} and separating the real and imaginary parts, we obtain the hydrodynamic form of the Schrödinger-Newton model. The imaginary part leads to the continuity equation
%\begin{equation}
%			\frac{\partial A}{\partial t} + \frac{1}{2m}  \left( 2 \nabla S \cdot \nabla A + A \nabla^2 S \right) = 0.
%		\end{equation}
\begin{equation}
\frac{\partial \rho}{\partial t} + \nabla \cdot (\rho \mathbf{u}) = 0, 
\label{ccc}
\end{equation}
while the real part leads to
		\begin{equation}
			\frac{\partial S}{\partial t} + \frac{1}{2m} (\nabla S)^2 
		+ \int A^2(\mathbf{r}') \Phi|\mathbf{r} - \mathbf{r}'| \, d\mathbf{r}' + Q = 0.
		\label{SN real}
		\end{equation}
		where
		\begin{equation}
			Q \equiv -\frac{\hbar^2}{2m} \frac{\nabla^2A}{A} = 
		-\frac{\hbar^2}{4m} \left( \frac{\nabla^2 \rho}{\rho} - \frac{1}{2} \frac{(\nabla \rho)^2}{\rho^2} \right)= -\frac{\hbar^2}{8 m}(\nabla \ln \rho)^2-\frac{\hbar^2}{4 m} \Delta(\ln \rho).
			\label{bohm}
		\end{equation}
is known the quantum potential, also known as the Bohm potential. Equation \eqref{SN real} is the quantum Hamilton-Jacobi (or Bernoulli) equation. By taking the gradient of this equation and dividing by $m$, we obtain the quantum counterpart of Euler equation (momentum-balance equation) 
\begin{equation}
\frac{\partial \mathbf{u}}{\partial t} + (\mathbf{u} \cdot \nabla) \mathbf{u} = -\frac{\nabla}{m} \int \rho(\mathbf{r}') \Phi|\mathbf{r} - \mathbf{r}'| \, d\mathbf{r}'- \frac{\nabla Q}{m}.
\label{Quantum euler}
\end{equation} 

Equations (\ref{ccc}) and (\ref{Quantum euler}) represent the quantum analogs of the classical hydrodynamic model (\ref{continuity})-(\ref{integrodifferential}). These can be treated in a manner analogous to the classical case discussed in Sec. \ref{Sec2}, namely, by linearizing the system and decomposing it into Fourier modes (see, for instance, Ref. \cite{31}). This procedure yields the following dispersion relation

\begin{equation}
\omega^2 = -\Omega_J^2 + \frac{\rho_0}{m} \widetilde{\Phi_c}(k) k^2 +\frac{\hbar^2}{4 m^2} k^4.
\label{Quantum dispersion}
\end{equation}

This is the quantum counterpart of the classical dispersion relation (\ref{dispersion final}), highlighting the interplay between generalized gravitational forces and quantum effects. It shows that quantum corrections introduce a term scaling as $k^4$ (in contrast to $k^2$ for classical pressure), which tends to stabilize the self-gravitating instability at large values of $k$, corresponding to small wavelengths.

By setting the correction potential $\Phi_c = 0$, the dispersion relation reduces, as expected, to

\begin{equation} \omega^2 = -\Omega_J^2 + \frac{\hbar^2}{4 m^2} k^4,
\end{equation}
which corresponds to the standard dispersion relation for a quantum self-gravitating system. In this case, the threshold between stability and instability, corresponding to $\omega=0$, is determined by the quantum Jeans wavenumber
\begin{equation} k_J = \left( \frac{2 m \Omega_J}{\hbar} \right)^{1/2}. 
\label{kjq}
\end{equation}
This applies, for instance, to boson stars or to Bose–Einstein condensate (BEC) dark matter models in the non-interacting (collisionless) regime \cite{Harko}. Interestingly, a hypothetical correction $\Phi_c$ scaling as $\sim 1/r^3$, which corresponds to a constant Fourier transform $\tilde{\Phi_c} = \text{const.}$, yields a dispersion relation (\ref{Quantum dispersion}) formally equivalent to that of BEC dark matter with self-interacting bosons, with $\text{const.} = {4\pi a \hbar^2 \rho_0}/{m^3}$, where $a$ is the s-wave scattering length \cite{Chavanis2}.

The inclusion of a correction to the Newtonian potential modifies this stability criterion for a quantum self-gravitating system within the framework of generalized gravity. In what follows, we examine how this modification manifests for the three types of potentials considered in this work: the Maneff potential, the logarithmic correction, and the Yukawa-type potential.

\subsection{Maneff Potential}
By substituting the corrective term of the Maneff potential into the dispersion relation \eqref{Quantum dispersion}, one has
\begin{equation}
\omega^2 = -\Omega_J^2 \left(1+  \frac{\pi }{2} B k \right) + \frac{\hbar^2}{4 m^2} k^4.
\label{Quantum dispersion maneff}
\end{equation}

Solving this equation for $k$ by setting $\omega=0$ leads to the critical wavenumber

\begin{equation}
k_J = \frac{1}{4} \left( \frac{\sqrt{2} \xi}{3^{1/3}} +
2 \sqrt{-\frac{\Delta^{1/3}}{2 \cdot 3^{2/3} d} + \frac{8 \Omega_J^2}{\left(3 \Delta\right)^{1/3}} +
\frac{\sqrt{2} \cdot 3^{1/3} \Omega_J^2 \pi B}{\xi d}} \right)
\label{Quantum k maneff}
\end{equation}
where we define $d = \frac{\hbar^2}{4m^2}$ and $\xi = \sqrt{\frac{-16 \cdot 3^{1/3} d \Omega_J^2 + \Delta^{2/3}}{d \Delta^{1/3}}}$ with

\begin{equation}
\Delta = 9 d \Omega_J^4 \pi^2 B^2 + \sqrt{3} \sqrt{d^2 \Omega_J^6 \left( 4096 d + 27 \Omega_J^2 \pi^4 B^4  \right)}.
\end{equation}

We can simplify this result for small $k$, the quantum pressure term $ \sim k^4$ can be neglected, leading to

\begin{equation}
\omega^2 = -\Omega_J^2 \left(1+  \frac{\pi }{2} B k \right).
\end{equation}
Since $B > 0$, the system exhibits instability for all perturbation modes, characterized by a purely imaginary frequency $\omega = i \omega_i$. The corresponding growth rate is given by

\begin{equation}
\omega_i = \sqrt{\Omega_J^2 \left(1+  \frac{\pi }{2} B k \right)}
\end{equation}
{indicating that, within this approximation, the instability becomes slightly more pronounced as $k$ increases, although the expression is only valid for small wavenumbers (long wavelengths).}

\subsection{Logarithmic Potential}
For the logarithmic potential, substituting the corrective term into the dispersion relation \eqref{Quantum dispersion} gives
\begin{equation}
\omega^2 = - \Omega_J^2 \left( 1-\frac{\pi}{2} \frac{\delta}{k}\right) + \frac{\hbar^2}{4 m^2} k^4.
\label{Quantum dispersion log}
\end{equation}
In this case, the critical wavenumber does not admit a simple closed-form solution. Nonetheless, valuable insights can still be gained by analyzing appropriate limiting cases. We consider two such limits: one for small values of $k$ (long wavelengths) and another for large values of $k$ (short wavelengths). Considering the small-$k$ limit where the $ \sim k^4$ term can be neglected, the dispersion relation simplifies to

\begin{equation}
\omega^2 = - \Omega_J^2 \left( 1-\frac{\pi}{2} \frac{\delta}{k}\right).
\end{equation}
%\textcolor{red}{For $\delta >0$, this leads to a critical wavenumber
%\begin{equation}
%\tilde{k_J} =\frac{\pi \delta}{2}.
%\end{equation}}
For $\delta < 0$, which is the most physically relevant case as discussed in the previous section, the system becomes unstable for all perturbation modes, since no effective pressure-like mechanism is present to counteract gravitational collapse. The corresponding growth rate is given by

\begin{equation}
    \omega_i = \sqrt{\Omega_J^2 \left( 1-\frac{\pi}{2} \frac{\delta}{k}\right)},
\end{equation}
which diverges in the limit {$k \to 0$ (long wavelengths)}. In the limit of large $k$, the $1/k$ contribution from the logarithmic correction becomes negligible, and the critical wavenumber asymptotically approaches the quantum Jeans wavenumber given by Eq. (\ref{kjq}).

%\begin{equation}
%\tilde{k_J} =\left(\frac{4 m^2 \Omega_J^2}{\hbar^2}\right)^{1/4}.
%\end{equation}

\subsection{Yukawa Potential}
For the Yukawa potential, the dispersion relation \eqref{Quantum dispersion} becomes
\begin{equation}
\omega^2 = -\Omega_J^2 \left(1+\frac{\alpha \lambda^2 k^2}{1 + \lambda^2 k^2}\right)  + \frac{\hbar^2}{4 m^2} k^4
\label{Quantum dispersion yukawa}
\end{equation}

By setting $\omega = 0$, we can solve for the critical wavenumber, yielding the following closed-form expression
\begin{equation}
    \tilde{k_J}^2= \frac{-2 + \frac{2 \cdot 2^{1/3} (d + 3 (1 + \alpha) \Omega_J^2 \lambda^4)}{\nabla} + \frac{2^{2/3} \nabla}{d}}{6\lambda^2},
    \label{Quantum k yukawa}
\end{equation}    
where we have defined
\begin{equation}
\nabla := \left( -2 d^3 - 9 (-2 + \alpha) \Omega_J^2 d^2 \lambda^4 + \sqrt{d^3 \left( d (2 d + 9 (-2 + \alpha) \Omega_J^2 \lambda^4)^2 - 4 (d + 3 (1 + \alpha) \Omega_J^2 \lambda^4)^3 \right)} \right)^{1/3}.
\end{equation}

For better physical insight, it is more useful to examine the limiting cases of the dispersion relation. For small $k$, neglecting the $ \sim k^4$, the dispersion relation (\ref{Quantum dispersion yukawa}) reduces to
%\begin{equation}
%\tilde{k^2_J} = \frac{-1}{ \lambda^2 \left(\alpha+1\right)}.
%\end{equation}
\begin{equation}
\omega^2 = -\Omega_J^2 \left(1+\frac{\alpha \lambda^2 k^2}{1 + \lambda^2 k^2}\right).
\end{equation}
For $\alpha >0$, the system is unstable for all perturbation modes, with the growth rate given by
\begin{equation}
\omega_i = \sqrt{\Omega_J^2 \left(1+\frac{\alpha \lambda^2 k^2}{1 + \lambda^2 k^2}\right)},
\end{equation}
which increases with $k$ and approaches $\omega_i = \Omega_J$ in the limit $k \to 0$ (infinite wavelength limit). For large $k$, approximating ${\lambda^2 k^2}/{1 + \lambda^2 k^2} \approx 1$, we obtain a critical wave number that is independent of $\lambda$, namely
\begin{equation}
\tilde{k_J}^2 =\left(\frac{4 m^2 \left(\alpha+1\right) \Omega_J^2}{\hbar^2}\right)^{1/2} = \sqrt{\alpha+1} k_J^2,
\end{equation}
which increases with the strength parameter $\alpha$ and diverges in the formal limit $\hbar \to 0$, as there is no pressure to counteract gravitational forces.
        		
\section{Conclusion}\label{Sec6}

In this paper, we investigated the Jeans instability in the context of modified gravitational theories, including both classical and quantum regimes. We adopted a hydrodynamic approach that is sufficiently general to accommodate arbitrary corrections to the Newtonian self-potential. We then focused on three specific examples of such corrections: Maneff-type, logarithmic, and Yukawa-type potentials. These cases encompass a broad range of possible deviations from Newtonian gravity, corresponding respectively to power-law, logarithmic, and exponential forms—functional behaviors that naturally arise in various theoretical frameworks or as weak-field limits of extended theories of gravity.

These different potentials have been proposed in the literature as simple yet effective modifications of Newtonian gravity for astrophysical applications. In this work, we tested their validity by comparing the modified stability criteria with observational data on Bok globules. These are small, dense molecular clouds, often considered the birthplaces of low-mass stars, and they provide an excellent laboratory for probing alternative gravitational theories, as their masses are close to the classical Jeans threshold. Our analysis demonstrated that each of the three corrections—Maneff-type, logarithmic, and Yukawa—can resolve the apparent discrepancies between the observed stability of Bok globules and the predictions of standard Newtonian theory, provided the model parameters lie within specific, constrained ranges. It was recently shown \cite{Our1} that the discrepancies observed in Bok globules can be effectively explained by invoking the presence of dark matter. In this context, the modified gravitational potentials considered here may be seen as phenomenological models that, to some extent, mimic certain gravitational effects typically attributed to dark matter.

A word of caution is in order, however. The parameter values required to match the observational data of Bok globules may differ by several orders of magnitude from those obtained in the literature based on other astrophysical observations. This discrepancy likely reflects the fact that the modified potentials considered here are not applicable across all scales and are not meant to serve as universal replacements for Newtonian gravity. Rather, they should be viewed as effective models designed to capture specific gravitational behaviors in particular physical contexts.

This paper suggests several potential directions for future research, both theoretical and observational. Theoretically, the derived generic dispersion relations might be a useful starting point for exploring other types of corrections that could arise from different theoretical frameworks. On the observational side, a larger and more varied dataset would be helpful in refining the accuracy of the results. Additionally, developing a more detailed collapse model for Bok globules, including factors such as turbulence and other dynamical effects, would offer the opportunity to refine the constraints and further validate these findings.

\appendix 
\section{}\label{A}
\begingroup

In the main text, we adopt a hydrodynamic approach to analyze gravitational instability. However, the problem can also be addressed from a kinetic perspective—using the Vlasov equation for a classical (collisionless) medium or the Wigner-Moyal formalism in the quantum regime. In this appendix, we outline the key steps of the kinetic treatment, focusing on the quantum case. The classical limit can be recovered by taking the formal limit $\hbar \to 0$.

The Wigner function $ W(\mathbf{r}, \mathbf{q}, t)$, associated with the wave function $\psi(\mathbf{r}, t)$, is defined as the Fourier transform of its autocorrelation function. That is,

\begin{equation}
W(\mathbf{r}, \mathbf{q}, t)=\int \psi^*(\mathbf{r}-\mathbf{s} / 2, t) \psi(\mathbf{r}+\mathbf{s} / 2, t) \exp (i\mathbf{q} \cdot \mathbf{s}) \mathrm{d} \mathbf{s}.
\end{equation}

This construction provides a phase-space representation of the quantum state, bridging the gap between quantum and classical descriptions. Using the standard Wigner–Moyal procedure \cite{Wigner,Moyal} and starting from the (generalized) Schrödinger-Newton equation (\ref{SN equation}), we can derive an evolution equation for the Wigner function, which is expressed as follows

\begin{equation}
\mathrm{i} \hbar\left(\frac{\partial}{\partial t}+\mathbf{v}_q \cdot \nabla\right) W= \int \tilde{\Phi}(\boldsymbol{\kappa}) \tilde{\rho } (\boldsymbol{\kappa}, t) \Delta W \mathrm{e}^{\mathrm{i} \boldsymbol{\kappa} \cdot \mathrm{r}} \frac{\mathrm{~d} \boldsymbol{\kappa}}{(2 \pi)^3},
\label{ap1}
\end{equation}
where $\mathbf{v}_q \equiv \hbar \mathbf{q}/m$ is the particle velocity and $\Delta W=\left[W^{-}-W^{+}\right]$, with $W^{ \pm}=W(\mathbf{r}, \mathbf{q} \pm \boldsymbol{\kappa} / 2, t)$. In Eq. (\ref{ap1}), we used the spectral density components

\begin{equation}
\tilde{\rho }  (\boldsymbol{\kappa}, t)=\int \rho (\mathbf{r}, t) \exp (-\mathrm{i} \boldsymbol{\kappa} \cdot \mathbf{r}) \mathrm{d} \mathbf{r},
\end{equation}
where the density is given by

\begin{equation}
\rho(\mathbf{r}, t)=|\psi(\mathbf{r}, t)|^2=\int W(\mathbf{r}, \mathbf{q}, t) \frac{\mathrm{d} \mathbf{q}}{(2 \pi)^3} .
\end{equation}

Equation (\ref{ap1}) can be analyzed in a manner similar to the quantum hydrodynamic model. We consider small perturbations, which are expressed as $W = W_0 + \tilde{W}$, where $W_0$ represents the equilibrium state and $\tilde{W}$ is the perturbed quantity. The perturbation is assumed to evolve in space and time as $\exp(i \mathbf{k} \cdot \mathbf{r} - i\omega t)$. By linearizing Eq. (\ref{ap1}) with respect to these perturbations, we can readily derive the dispersion relation (see e.g. \cite{5,Tito} for mathematical details)
\begin{equation}
1- \tilde{\Phi} (\mathbf{k}) \int \frac{\Delta W_0}{\hbar\left(\omega-\mathbf{k} \cdot \mathbf{v}_q\right)} \frac{\mathrm{d} \mathbf{q}}{(2 \pi)^3}=0 .
\label{ap2}
\end{equation}

If we consider the zero-temperature limit, where the distribution collapses to a Dirac delta function, i.e., $W_0(\mathbf{q})=(2 \pi)^3 \rho_0 \delta\left(\mathbf{q}-\mathbf{q}_0\right)$, Equation (\ref{ap2}) reduces to the dispersion relation (\ref{Quantum dispersion}) derived within the quantum hydrodynamic approach. The kinetic approach, however, allows us to go further by accounting for kinetic effects, including finite temperature corrections. Assuming an isotropic distribution characteristic of nearly equilibrium conditions, and considering small thermal corrections, Equation (\ref{ap2}) reduces to (detailed computations can be found e.g. in \cite{5,Tito})

\begin{equation}
\omega^2 = -\Omega_J^2 + \frac{\rho_0}{m} \widetilde{\Phi_c}(k) k^2 +  \langle v^2 \rangle k^2+\frac{\hbar^2}{4 m^2} k^4.
\end{equation}

This expression smoothly interpolates between the purely quantum dispersion relation (\ref{Quantum dispersion}) and the classical dispersion relation (\ref{dispersion final}), including thermal pressure corrections, upon identifying $c_s^2 \equiv \langle v^2 \rangle$. For long wavelengths, the quantum correction term becomes negligible, and the classical dispersion relation (\ref{dispersion final}) is recovered. At the opposite limit, for short wavelengths, the quantum correction term, scaling as $\sim k^4$, dominates over the classical pressure, restoring Eq. (\ref{Quantum dispersion}).


\begin{thebibliography}{50}
\bibitem{1} C.M. Will, Theory and Experiment in Gravitational Physics, 2nd ed. (Cambridge University Press, 2018). \url{https://doi.org/10.1017/9781316338612}
\bibitem{2} C.M. Will, The Confrontation between General Relativity and Experiment, \href{https://doi.org/10.12942/lrr-2014-4}{Living Rev. Relativ. \textbf{17}, 4 (2014).}  
\bibitem{3} F. Zwicky, Republication of: The redshift of extragalactic nebulae, \href{https://doi.org/10.1007/s10714-008-0707-4}{Gen. Relativ. Gravit. \textbf{41} , 207 (2009).} 




\bibitem{p1} A. De Felice, S. Tsujikawa, $f(R)$ theories, \href{https://doi.org/10.12942/lrr-2010-3}{Living Rev. Rel. \textbf{13}, 3 (2010).}   

\bibitem{p2} C.H. Brans, R.H. Dicke, Mach’s principle and a relativistic theory of gravitation, \href{https://doi.org/10.1103/PhysRev.124.925}{Phys. Rev. \textbf{124}, 925 (1961).} 

\bibitem{p3} P.G. Bergmann, Comments on the scalar–tensor theory, \href{https://doi.org/10.1007/BF00668828}{Int. J. Theor. Phys. \textbf{1}, 25 (1968).}  

\bibitem{p4} J.B. Jimenez, L. Heisenberg, T.S. Koivisto, The geometrical trinity of gravity, \href{https://doi.org/10.3390/universe5070173}{Universe \textbf{5}, 173 (2019).}

\bibitem{p5} M.P. Dabrowski, K. Marosek, Regularizing cosmological singular-ities by varying physical constants, \href{https://doi.org/10.1088/1475-7516/2013/02/012}{JCAP \textbf{1302}, 012 (2013).}   

\bibitem{p6} K. Leszczynska, A. Balcerzak, M.P. Dabrowski, Varying constants quantum cosmology, \href{https://doi.org/10.1088/1475-7516/2015/02/012}{JCAP \textbf{1502}, 012 (2015).}  

\bibitem{p7} T.P. Sotiriou, V. Faraoni, $f(R)$ theories of gravity, \href{https://doi.org/10.1103/RevModPhys.82.451}{Rev. Mod. Phys. \textbf{82}, 451 (2010).}   

\bibitem{p8} S. Capozziello, M. De Laurentis, Extended theories of gravity, \href{https://doi.org/10.1016/j.physrep.2011.09.003}{Phys. Rep. \textbf{509}, 167 (2011). } 

\bibitem{Milgrom1} M. Milgrom, A modification of the Newtonian dynamics as a possible alternative to the hidden mass hypothesis, \href{https://ui.adsabs.harvard.edu/abs/1983ApJ...270..365M/abstract}{Astrophys. J. \textbf{270}, 365 (1983).} 
\bibitem{Maneff} G. Maneff, La gravitation et le principe de l'égalité de l'action et de la réaction, Comptes Rendus Acad. Sci. Paris \textbf{178}, 2159 (1924).


\bibitem{Maneff2} G. Maneff, Le principe de la moindre action et la gravitation, Comptes Rendus Acad. Sci. Paris \textbf{190}, 963 (1930).

\bibitem{Maneff3} G. Maneff, La gravitation et l'énergie au zero, Comptes Rendus Acad. Sci. Paris \textbf{190}, 1374 (1930). 

\bibitem{Tohline} J. E. Tohline, The Internal Kinematics and Dynamics of Galaxies, IAU Symp. 10; E. Athanassoula, Ed.; Reidel: Dordrecht, The Netherlands, 1983; pp. 205–206.

\bibitem{Kuhn} J. R. Kuhn and L. Kruglyak, Non-Newtonian forces and the invisible mass problem, \href{https://doi.org/10.1086/164942}{Astrophys. J. \textbf{313}, 1 (1987).}

\bibitem{Smullin} S. J. Smullin \textit{et al.}, Constraints on Yukawa-type deviations from Newtonian gravity at 20 microns, \href{https://doi.org/10.1103/PhysRevD.72.122001}{Phys. Rev. D \textbf{72}, 122001 (2005).}


\bibitem{Nonlocal} T. Biswas, E. Gerwick, T. Koivisto, and A. Mazumdar,
Towards Singularity- and Ghost-Free Theories of Gravity, \href{https://doi.org/10.1103/PhysRevLett.108.031101}{Phys. Rev. Lett. \textbf{108}, 031101 (2012).}

\bibitem{Blome} J.-H. Blome, C. Chicone, F. W. Hehl, and B. Mashhoon, Nonlocal modification of Newtonian gravity, \href{https://doi.org/10.1103/PhysRevD.81.065020}{Phys. Rev. D, \textbf{81}, 065020
(2010).}

\bibitem{Blome2} {F. W. Hehl and B. Mashhoon, Nonlocal gravity simulates dark matter,} \href{https://doi.org/10.1016/j.physletb.2009.02.033}{Phys. Lett. B \textbf{673}, 279 (2009).}

\bibitem{Ourabah} K. Ourabah, The other way around: from alternative gravity to entropy, \href{https://doi.org/10.1088/1361-6382/ad0eeb}{Class. Quantum Grav. \textbf{41}, 015010 (2024).}

\bibitem{J1} I. De Martino, A. Capolupo, Kinetic theory of Jean instability in Eddington-inspired Born–Infeld gravity, \href{https://doi.org/10.1140/epjc/s10052-017-5300-0}{Eur. Phys. J. \textbf{77}, 715(2017).} 



\bibitem{J2} A. Bessiri, K. Ourabah, T.H. Zerguini, Jeans instability in Eddington-inspired Born–Infeld (EiBI) gravity: a quantum approach, \href{http://doi.org/10.1088/1402-4896/ac1cd2}{Phys. Scr. \textbf{96}, 125208 (2021).} 

\bibitem{J3} S. Capozziello, M. De Laurentis, I. De Martino, M. Formisano, and S.D. Odintsov, Jeans analysis of self-gravitating systems in $f(R)$ gravity, \href{https://doi.org/10.1103/PhysRevD.85.044022}{Phys. Rev. D \textbf{85}, 044022 (2012).} 

\bibitem{J4} M. Roshan, S. Abbassi, Jeans analysis in modified gravity, \href{https://doi.org/10.1103/PhysRevD.90.044010}{Phys.Rev. D \textbf{90}, 044010 (2014)}  


\bibitem{J5} C. Gomes, Jeans instability in non-minimal matter-curvature coupling gravity, \href{https://doi.org/10.1140/epjc/s10052-020-8189-y}{Eur. Phys. J. C \textbf{80}, 633 (2020).} 

\bibitem{J6} C. Gomes and K. Ourabah, Quantum kinetic theory of Jeans instability in non-minimal matter-curvature coupling gravity, \href{https://doi.org/10.1140/epjc/s10052-023-11184-9}{Eur. Phys. J. C \textbf{83}, 40 (2023).}  

\bibitem{J7} G.M. Kremer, Jeans instability from post-Newtonian Boltzmann equation, \href{https://doi.org/10.1140/epjc/s10052-021-09728-y}{Eur. Phys. J. C \textbf{81}, 927 (2021).}

\bibitem{J8} G.M. Kremer and K. Ourabah, A self-gravitating system composed of baryonic and dark matter analysed from the post-Newtonian Boltzmann equations, \href{https://doi.org/10.1140/epjc/s10052-023-12000-0}{Eur. Phys. J. C \textbf{83}, 819 (2023).} 











\bibitem{4} S. Kenyon and S. Starrfield, On the structure of BOK globules, \href{https://doi.org/10.1086/130483}{PASP 91, 271 (1979).}  

\bibitem{5} K. Ourabah, Collective Phenomena in Plasmas and Elsewhere: Kinetic and Hydrodynamic Approaches, 1st ed. (Wiley, 2023). \url{https://doi.org/10.1002/9781394236756}
\bibitem{6} K. Ourabah, Jeans analysis in fractional gravity, \href{https://doi.org/10.1140/epjc/s10052-024-13443-9}{Eur. Phys. J. C \textbf{84}, 1047 (2024).} 













\bibitem{H1} M. Kiessling, The “Jeans swindle”: A true story—mathematically speaking, \href{https://www.sciencedirect.com/science/article/pii/S0196885802005560}{Adv. Appl. Math. \textbf{31}, 132 (2003).} 

\bibitem{H2} M. Joyce, B. Marcos, and F. S. Labini, Dynamics of finite and infinite self-gravitating systems with cold quasi-uniform initial conditions, \href{https://iopscience.iop.org/article/10.1088/1742-5468/2009/04/P04019}{J. Stat. Mech. P04019 (2009).}  

\bibitem{H3} J. Peebles, \textit{Large-Scale Structures of the Universe} (Princeton University Press, 1980). 


\bibitem{H4} P.-H. Chavanis, Gravitational instability of finite isothermal spheres, \href{	https://doi.org/10.1051/0004-6361:20011424}{Astron. Astrophys. \textbf{381}, 340 (2002).}


\bibitem{Bonnor} W. B. Bonnor, Jeans' formula for gravitational instability, \href{https://articles.adsabs.harvard.edu//full/1957MNRAS.117..104B/0000106.000.html}{Mon. Not. R. Astron.
Soc. \textbf{117}, 104 (1957).} 



\bibitem{7} G. Maneff, Die Masse der Feldenergie und die Gravitation, Astron Nachr 236, 401 (1929). \href{https://doi.org/10.1002/asna.19292362402}{Astron. Nachr. \textbf{236}, 401 (1929).}
\bibitem{8} F. Diacu, V. Mioc, and C. Stoica, Phase-space structure and regularization of Manev-type problems, Nonlinear Analysis: Theory, Methods  Applications 41, 1029 (2000). \href{https://doi.org/10.1016/S0362-546X(98)00326-5}{Nonlinear Anal. \textbf{41}, 1029 (2000).}
\bibitem{9} R. Ivanov and E. Prodanov, Manev Potential and General Relativity, (2005). \url{https://doi.org/10.21427/saqk-qf53} 
\bibitem{10} S. Kirk, I. Haranas, and I. Gkigkitzis, Satellite motion in a Manev potential with drag, \href{https://doi.org/10.1007/s10509-012-1330-0}{Astrophys. Space Sci. \textbf{344}, 313 (2013).}  

\bibitem{Man} F. S. Escórcio, J. C. Fabris, J. D. Toniato, and H. Velten, Celestial mechanics in Newtonian-like gravity with variable $G$, \href{https://doi.org/10.1140/epjp/s13360-023-04729-0}{Eur. Phys. J. Plus \textbf{138}, 1084 (2023).} 

\bibitem{Sheykhi} A. Sheykhi and S. H. Hendi, Power-law entropic corrections to Newton’s law and Friedmann equations
, \href{https://doi.org/10.1103/PhysRevD.84.044023}{Phys. Rev. D \textbf{84}, 044023 (2011).}

\bibitem{12} J.C. Fabris and J. P. Campos, Spiral galaxies rotation curves with a logarithmic corrected newtonian gravitational potential, \href{https://doi.org/10.1007/s10714-008-0654-0}{Gen. Relativ. Gravit. \textbf{41}, 93 (2009).}  

\bibitem{13} W. H. Kinney and M. Brisudova, An Attempt to Do Without Dark Matter, \href{https://doi.org/10.1111/j.1749-6632.2001.tb05627.x}{Ann. N. Y. Acad. Sci. \textbf{927}, 127 (2001).}


\bibitem{14} A. A. Kirillov, The nature of dark matter, \href{https://doi.org/10.1016/j.physletb.2005.11.005}{Physics Letters B \textbf{632}, 453 (2006).}  



\bibitem{16} C. Lu, Z.-W. Li, S.-F. Yuan, Z. Wan, S.-H. Qin, K. Zhu, and Y. Xie, Preliminary limits of a logarithmic correction to the Newtonian gravitational potential in binary pulsars, \href{https://doi.org/10.1088/1674-4527/14/10/009}{Res. Astron. Astrophys. \textbf{14}, 1301 (2014).}

\bibitem{Deng} Deng, XM., Xie, Y. Preliminary limits on a logarithmic correction to the Newtonian gravitational potential in the solar system, \href{https://doi.org/10.1007/s10509-013-1735-4}{Astrophys. Space Sci \textbf{350}, 103 (2014)}

\bibitem{m} M. J. Lighthill, \textit{Introduction to Fourier Analysis and Generalised Functions}, Cambridge University Press, 1958.

\bibitem{17} A. Giusti, MOND-like fractional Laplacian theory, \href{https://doi.org/10.1103/PhysRevD.101.124029}{Phys. Rev. D \textbf{101}, 124029 (2020).}  


\bibitem{18} H. Yukawa, On the Interaction of Elementary Particles. I, \href{https://doi.org/10.1143/PTPS.1.1}{Prog. Theor. Phys. Suppl. \textbf{1}, 1 (1955).}
\bibitem{19} R. T. Farouki and S. Hamaguchi, Thermodynamics of strongly-coupled Yukawa systems near the one-component-plasma limit. II. Molecular dynamics simulations, \href{https://doi.org/10.1063/1.467955}{J. Chem. Phys. \textbf{101}, 9885 (1994).}
\bibitem{20} D. J. Kapner, T. S. Cook, E. G. Adelberger, J. H. Gundlach, B. R. Heckel, C. D. Hoyle, and H. E. Swanson, Tests of the Gravitational Inverse-Square Law below the Dark-Energy Length Scale, \href{https://doi.org/10.1103/PhysRevLett.98.021101}{Phys. Rev. Lett. \textbf{98}, 021101 (2007).}


\bibitem{21} L. Iorio, Constraints on the range $\lambda$ of Yukawa-like modifications to the Newtonian inverse-square law of gravitation from Solar System planetary motions, \href{https://doi.org/10.1088/1126-6708/2007/10/041}{J. High Energy Phys. \textbf{2007}, 041 (2007).} 


\bibitem{22} P. Jovanović, V. Borka Jovanović, D. Borka, and A. F. Zakharov, Constraints on Yukawa gravity parameters from observations of bright stars, \href{https://doi.org/10.1088/1475-7516/2023/03/056}{J. Cosmol. Astropart. Phys. \textbf{2023}, 056 (2023).}  


\bibitem{23} I. Haranas and I. Gkigkitzis, Basic thermodynamic parameters of a black hole resulting from a Yukawa type of correction to the metric, \href{https://doi.org/10.1007/s10509-011-0865-9}{Astrophys. Space Sci. \textbf{337}, 693 (2012).} 


\bibitem{24} R. Kandori et al., Near-Infrared Imaging Survey of Bok Globules: Density Structure, \href{https://doi.org/10.1086/444619}{Astrophys. J. \textbf{130}, 2166 (2005).} 


\bibitem{25} J. Vainio and I. Vilja, Jeans analysis of Bok globules in $f(R)$ gravity, \href{https://doi.org/10.1007/s10714-016-2120-8}{Gen. Relativ. Gravit. \textbf{48}, 129 (2016). } 


\bibitem{Our1} K. Ourabah, Gravitational instability with a dark matter background: exploring the different scenarios, \href{https://doi.org/10.1140/epjc/s10052-022-10529-0}{Eur. Phys. J. C \textbf{82}, 565 (2022).}

%\bibitem{Our2} G. M. Kremer and K Ourabah, A self-gravitating system composed of baryonic and dark matter analysed from the post-Newtonian Boltzmann equations, \href{https://doi.org/10.1140/epjc/s10052-023-12000-0}{Eur. Phys. J. C \textbf{83}, 819 (2023).}  


\bibitem{M1} H. Moradpour, A. H. Ziaie, S. Ghaffari, F. Feleppa, The generalized and extended uncertainty principles and their implications on the Jeans mass, \href{https://doi.org/10.1093/mnrasl/slz098}{MNRAS Lett. \textbf{488}, L39 (2019).}

\bibitem{M2} H. Shababi and K. Ourabah, Non-Gaussian statistics from the generalized uncertainty principle, \href{https://doi.org/10.1140/epjp/s13360-020-00726-9}{Eur. Phys. J. Plus \textbf{135}, 697 (2020).} 

\bibitem{26} I. De Martino, R. Lazkoz, and M. De Laurentis, Analysis of the Yukawa gravitational potential in $f(R)$ gravity. I. Semiclassical periastron advance, \href{https://doi.org/10.1103/PhysRevD.97.104067}{Phys. Rev. D \textbf{97}, 104067 (2018).}  

\bibitem{27} R. Ruffini and S. Bonazzola, Systems of Self-Gravitating Particles in General Relativity and the Concept of an Equation of State, \href{https://doi.org/10.1103/PhysRev.187.1767}{Phys. Rev. \textbf{187}, 1767 (1969).}  

\bibitem{28} L. Diósi, Gravitation and quantum-mechanical localization of macro-objects, \href{https://doi.org/10.1016/0375-9601(84)90397-9}{Phys. Lett. A \textbf{105}, 199 (1984).} 


\bibitem{29} R. Penrose, On Gravity’s role in Quantum State Reduction, \href{https://doi.org/10.1007/BF02105068}{Gen. Relat. Gravit. \textbf{28}, 581 (1996).}  


\bibitem{30} S.A. Khan and M. Bonitz, Quantum Hydrodynamics, in Complex Plasmas, edited by M. Bonitz, J. Lopez, K. Becker, and H. Thomsen, \href{https://doi.org/10.1007/978-3-319-05437-7_4}{Springer Int. Publ. \textbf{82}, 103 (2014).}

\bibitem{32} E. Madelung, Quantentheorie in hydrodynamischer Form, \href{https://doi.org/10.1007/BF01400372}{Z. Physik \textbf{40}, 322 (1927).} 


\bibitem{31} K. Ourabah, On the collective properties of quantum media, \href{https://doi.org/10.1140/epjp/s13360-022-03641-3}{Eur. Phys. J. Plus \textbf{138}, 55 (2023).} 


\bibitem{Harko} C. G. Böhmer, T. Harko, Can dark matter be a Bose–Einstein con-densate?, \href{https://doi.org/10.1088/1475-7516/2007/06/025}{J. Cosmol. Astropart. Phys. \textbf{6}, 025 (2007).} 

\bibitem{Chavanis2} P. H. Chavanis, Jeans Instability of Dissipative Self-Gravitating
Bose–Einstein Condensates with Repulsive or
Attractive Self-Interaction: Application to
Dark Matter, \href{https://doi.org/10.3390/universe6120226}{Universe \textbf{6}, 226 (2020).}

\bibitem{Wigner} E. P. Wigner, {On the Quantum Correction For Thermodynamic Equilibrium}, \href{https://doi.org/10.1103/PhysRev.40.749}{Phys. Rev. \textbf{40}, 749 (1932).}
\bibitem{Moyal} J. E. Moyal, Quantum mechanics as a statistical theory, in \textit{Mathematical Proceedings of the Cambridge Philosophical
Society} (Cambridge University Press, Cambridge, 1949), Vol.
45, pp. 99–124.

\bibitem{Tito} J. T. Mendonça and H. Terças, \textit{Physics of Ultra-Cold
Matter}, Springer Series on Atomic, Optical and Plasma
Physics \textbf{70} (2013).

\end{thebibliography}
\end{document}